\documentclass[%
 reprint,
 amsmath,amssymb,
 aps, pre
]{revtex4-2}

\usepackage{graphicx}
\usepackage{dcolumn}
\usepackage{bm}
\usepackage{hyperref}
\usepackage[mathlines]{lineno}

\hypersetup{
    colorlinks=true,
    linkcolor=blue,
    filecolor=magenta,      
    urlcolor=cyan,
}

\newcommand{\dist}{\sim}

\newcommand{\N}{\mathcal{N}}

\newcommand{\FN}{\text{FN}}
\newcommand{\CN}{\text{CN}}
\newcommand{\DFN}{\text{DFN}}

\newcommand{\PD}{\text{PD}}

\begin{document}

\preprint{APS/123-QED}

\title{Stochastic methods for slip prediction in a sheared granular system}

\author{P.~Bretz}
 \email{pbretz@ou.edu}
\affiliation{Department of Mathematics, University of Oklahoma, Norman, Oklahoma 73019, USA\\}
\author{L.~Kondic}
    \email{kondic@njit.edu}
\affiliation{Department of Mathematical Sciences, New Jersey Institute of Technology, Newark, New Jersey 07102, USA\\}
\author{M.~Kramar}
 \email{miro@ou.edu}
\affiliation{Department of Mathematics, University of Oklahoma, Norman, Oklahoma 73019, USA}

\date{\today}

\begin{abstract}
We consider a sheared granular system experiencing intermittent dynamics of stick-slip type via discrete element simulations.  The considered setup consists of a two-dimensional system of soft frictional particles sandwiched between solid walls, one of which is exposed to a shearing force.  The slip events are detected using stochastic state space models applied to various measures describing the system.  We show that the measures describing the forces between the particles  provide earlier detection of an upcoming slip event than the measures based solely on the wall movement. By comparing the detection times obtained from the considered measures, we observe that a typical  slip event starts  with a local change in the force network.  However, some local changes do not  spread globally over the force network.  For the changes that become global, we find a sharp critical value for their size. If the size of a global change exceeds the critical  value, then it triggers a slip event; if it does not, then a much weaker micro-slip follows. Quantification of the changes in the force network is made possible by formulating clear and precise measures describing their static and dynamic properties. \end{abstract}

\maketitle

\section{\label{sec:introduction}Introduction}

Stick-slip dynamics is ubiquitous in granular and soft matter systems as well as in many other ones, see~\cite{arcangelis_physrep_2016,falk_langer_2011} for reviews. Understanding the intermittent stick-slip dynamics is of utmost importance because of the relevance to large-scale possibly cataclysmic events such as earthquakes.  Therefore, there is a large body of research focusing on quantifying and explaining intermittency in a variety of systems, see~\cite{daub_carlson_2010,kawamura_2012,arcangelis_physrep_2016,luding2021jamming} for reviews. For obvious reasons, there is an active interest in  identifying precursors to the slip events that lead to abrupt structural rearrangements of the systems, see~\cite{scuderi_natgeoscience_2016}  for a recent example in the context of earthquakes. Extensive research  considering acoustic precursors of intermittent dynamics in laboratory experiments can be found in~\cite{johnson_geophys_2013} and references therein.

The attempts to identify precursors of slip events, fractures, and shear bands have a long history; for brevity here we discuss only recent developments.  Several precursors are identified in various experimental systems~\cite{nerone_pre_2003,aguirre_pre_2006,scheller2006precursors,zaitsev_epl_2008}. These works report the existence of structural rearrangements of particles and point out that they become more prominent as a slip event approaches.  Consistent results are also obtained in simulations~\cite{staron2002preavalanche,Maloney2004_prl,ciamarra_prl10,welker2011precursors,dahmen_natcom_2016,dahmen_group15,Denisov2017_sr,bares_pre17}. 
However, despite a large amount of work concerned with identifying the precursors of slip events, we are not aware that they have been used for predictive purposes so far. 

\begin{figure}
    \centering
    \includegraphics[width=0.48\textwidth]{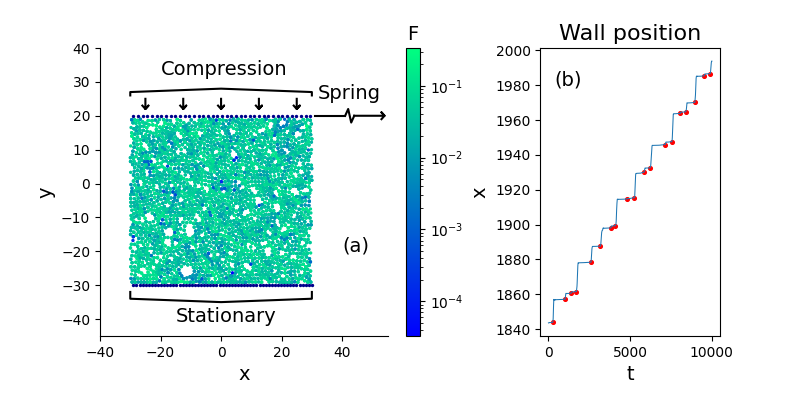}
    \caption{(a) Snapshot of simulated particles at a fixed time, with the magnitude of the normal force on each particle shown on the logarithmic scale. The top and bottom walls are rigid rows of particles; the bottom wall is fixed and the top wall is exposed to normal (-$y$ direction) compression and to horizontal (+$x$ direction) elastic force. (b) The $x$ position of the top wall for the first 10,000 time outputs. The red dots mark  the start  of  slip events determined by the offline method described in \ref{sec:offline_method}. 
    }
    \label{fig:simulation_example}
\end{figure}

In our recent work~\cite{kramar2022intermittency}, we considered a two dimensional system of soft disk-like particles simultaneously exposed  to compression and shear, see Fig.~\ref{fig:simulation_example}(a). If  the pressure is sufficiently large and the shearing speed is sufficiently small, then this system is found to exhibit  intermittent dynamics of the stick-slip type~\cite{ciamarra_prl10}.  By analyzing the outcome of discrete element simulations (DEMs), in~\cite{kramar2022intermittency} we demonstrate  a fundamental difference between the measures  defined as system-wide averages, and  the measures that quantify the evolution of the system on a micro (particle) or mesoscale scale. The system-wide averages are found to be almost insensitive to an approaching slip, while the latter  measures show a promising  potential to serve as precursors of slip events, and will be  considered in the present paper.

One significant obstacle in using the  measures quantifying the evolution of the system on a micro or mesoscale scale for predicting an approaching slip event is that they are rather noisy and exhibit  considerable variations between individual slip events. The method for delineating the slip events introduced  in~\cite{kramar2022intermittency} is based on the wall velocity and requires the knowledge of the whole data set.  Therefore,  to identify the start of a slip event  at time $t_0$, the data for both $t\leq t_0$ and $t> t_0$ are used.  We will refer to this method as the `offline method' and note that  such an approach, while logical and commonly used, is not suitable for predictive purposes.

In the present paper, we develop a  method that can predict the slip events on-the-fly for an incoming stream of data. We will refer to this method as the `online method.' The online method implemented in this paper is based on stochastic state space models (SSSMs)~\cite{petris2009dlm} that are often used to model an observable signal by decomposing it into several dynamic components, such as trend (steadily increasing/decreasing part of the signal), seasonal (oscillations of the signal with different frequencies) or regressive components. Typically, these components are  not directly observable but they can be modeled and used to predict the future values of the signal. SSSMs have been successfully used in meteorology~\cite{evensen2009da}, robotics to guide autonomous vehicles~\cite{kwon2013kalmanrobotics}, and track submarines~\cite{stone2014bayesiantracking} as well as in economics~\cite{zhao2015kalmanmri, shankar2012kalmaneconomics}.  A closely related application to our problem can be found in~\cite{tordesillas2020spatial} where the authors consider stochastic methods for modeling high-dimensional non-stationary spatio-temporal data collected at the sites with geological hazards. Another related application is the usage of machine learning to identify spatio-temporal clusters of surface displacement that are indicative of an imminent rock slide~\cite{tordesillas_remote_2019}.

The applicability of SSSMs stems from the fact that the behavior of the considered measures is fundamentally different during the stick and slip regimes. The  SSSMs allow us to predict the next value of a considered measure with good accuracy as long as the system is in the stick regime. The model is designed in such a way that it is too rigid to adapt to the evolution of the measure during the slip regime and its predictions become inaccurate when the system enters this regime.  Thus, to detect  the start of a slip event, we analyze the differences between the values predicted by the SSSM, and the observed  values. In particular, we identify the start of a slip event as the time at which the error in our predictions becomes larger than a  maximal acceptable value of the error, $T_e$.  As we will see, the online method based on SSSMs is rather general and can be in principle  applied to any scalar measure that describes the evolution. Naturally, the upcoming slip event can be detected in advance only if the behavior of the considered measure changes before the slip event starts.  For this reason, it is important to identify appropriate measures to consider. Here, we are guided by our previous work~\cite{kramar2022intermittency}, which shows the predictive potential of the measures  based on force networks.  We will show in particular that the measures that are sensitive to {\em local changes} in the force network provide appropriate input to SSSMs.  

One important part of the present paper is the investigation of how the detection time of the slip events depends on the considered measure and the value of  $T_e$. As the value of $T_e$ decreases, the upcoming slip events are detected earlier but the number of false positives increases. We will see that these false positives correspond to the so-called micro-slips which were reported in previous work on granular systems, see e.g.~\cite{Dalton2001_pre,Johnson:2005kt,Johnson:2007cs,ciamarra_prl10,long_granmat_2019}, or local changes of the force network and possibly particle positions that do not lead to wall motion, see also~\cite{bares_pre17}.  By using SSSMs, we find that there is a clear separation between the  slip events and micro-slips, similarly observed in previous works~\cite{arcangelis_iop11,long_granmat_2019}. 

We also show  that both slip events and micro-slip events  are initiated by a local change of the force network, which first becomes global by spreading over the network, and only afterward does the shearing wall start moving. Some local changes dissipate instead of spreading and therefore do not lead to slip or micro-slip  events. This connection between local and global changes and resulting wall activity has been discussed in the literature, see, e.g.~\cite{Maloney2004_prl,dauchot05,bares_pre17,manning_sm22}; however, the advancement in quantification of force networks and data analysis implemented in the present work now allows that connection to be explored for predictive purposes.

The rest of this paper is organized as follows. Section~\ref{sec:simulations} describes the simulated system, and the measures used for predictions are presented in Sec.~\ref{sec:measures}. In Sec.~\ref{sec:detection} we introduce the framework for model-based slip detection. The general framework for stochastic models and the Kalman  filter are reviewed in Sec~\ref{sec:kalman},  while the models for the considered measures are specified in Sec.~\ref{sec:models}. The main results are presented in Sec.~\ref{sec:results}, followed by the conclusions in Sec.~\ref{sec:conclusions}.

\section{Methods}
\label{sec:methods}

\subsection{Discrete element simulations} 
\label{sec:simulations}

The simulations of dense granular material are carried out within a setup that leads to the stick-slip dynamics~\cite{kramar2022intermittency}. The simulation techniques are identical to the ones described in detail in the previous  work~\cite{kovalcinova_scaling} and here we provide just a summary.  The material parameters used are motivated by experiments with photoelastic  particles described e.g.~in~\cite{dijksman_2018}.
Figure~\ref{fig:simulation_example} illustrates the simulated system and its behavior.   The granular particles are modeled as 2D soft frictional disks. We place $2500$ disks (system particles) between two horizontal rough walls placed parallel to the $x$ axis,  see Fig.~\ref{fig:simulation_example}(a).  The system particles are bi-disperse, with 25\% of large particles and 75\% of small particles with a size ratio of 1.25:1.  The walls are constructed from small particles which are at a fixed distance from each other, see~\cite{kramar2022intermittency}. The bottom wall is kept fixed,  while the top one is pulled by a harmonic spring moving with the velocity $v_{spring}$ in the $+x$ direction.
 
The linear spring-dashpot model is used to describe the interactions between the system particles and between the system and wall particles. We use the diameter of small particles, $d$, as the length scale, their mass, $m$, as the mass scale, and the binary collision time, $\tau_c$, as the time scale. Motivated by experiments ~\cite{dijksman_2018}, we use $d=1.27$ cm, $m=1.32$ g, and $\tau_c = 1.25\times10^{-3}$ s, as appropriate for  particles of Young modulus of $Y\approx 0.7$ MPa. Then, the normal spring constant is $k_n = m\pi^2/2{\tau_c}^2 \approx 4.17$ N/m,  and the tangential spring constant (needed for modeling of tangential forces using the Cundall-Strack model~\cite{cundall79}) is taken as $k_t = 6k_n/7$. The coefficient of static friction, $\mu$, is equal to $0.7$ for particle-particle contacts and $\mu = 2$ for particle-wall contacts (using a larger value here to reduce slipping of the system particles relative to the walls).  The force constant of the spring applied to the top wall, $k_s$, is significantly smaller than the one describing particle interactions, $k_s = k_n/400$. The (constant) restitution coefficient is 0.5. In addition, a normal compression force is applied in the 
$-y$ direction to  model an externally applied pressure (force/length in 2D) of $p=0.02$; gravitational effects are not included.  We note that with our choice of units, the numerical value of the applied pressure is of the same order of magnitude as the average overlap (compression) of the particles.  

Stick-slip dynamics occurs for sufficiently large applied pressure and for sufficiently slow shearing, see, e.g.~\cite{ciamarra_prl10}. For $p = 0.02$ we find by experimenting that the value $v_{spring} =  1.5\times10^{-3}$ leads to stick-slip.  Then, we integrate Newton's equations of motion for both the translation and rotational degrees of freedom using a fourth-order predictor-corrector method with time step $dt = 0.02$. The states of the system, used to compute the quantities presented in this paper,  are stored at times $\delta t = 10 dt$ apart.  In this paper, all the figures that show time-dependent results are in units of $\delta t$.

The simulations start by applying a pressure $p$ to the top wall and then letting the system 
relax until the ratio of kinetic/potential energy becomes sufficiently 
small~\cite{kramar2022intermittency}.  Then we start moving the spring in the $+x$ direction, and, after the initial transient regime disappears, start collecting data.  Figure~\ref{fig:simulation_example}(b) shows the horizontal position of the top wall  for a short time window; the red dots illustrate the starts of slip events obtained using the offline method summarized in Sec.~\ref{sec:offline_method}. The complete simulation contains  $5\times 10^5$ time output (first $10^4$ outputs are shown in Fig.~\ref{fig:simulation_example}(b)) and  hundreds of slip events.

\subsection{Measures}
\label{sec:measures}

\begin{figure}
    \centering
    \includegraphics[width=0.48\textwidth]{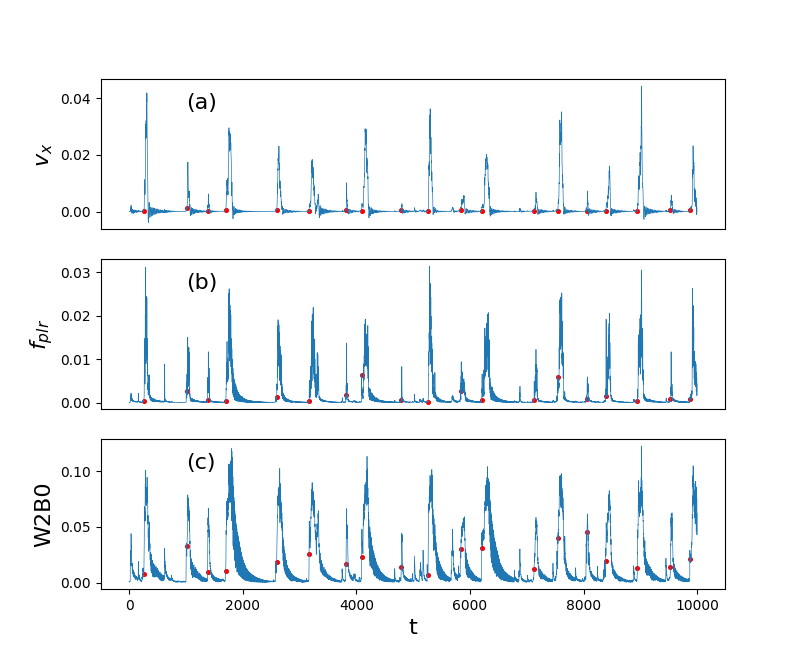}
    \caption{Sample of the evolution for the first 10,000 outputs: (a) the magnitude of the velocity, $v_x$, of the top wall,
    (b) the left-right percolation force, $f_{plr}$, and (c) the Wasserstein distance, W2B0. All the quantities are defined in Sec.~\ref{sec:measures}.  The red dots mark  the start  of  slip events determined by the offline method described in Sec.~\ref{sec:offline_method}. Here, $v_x$ shows two alternating regimes of behavior: long periods of stick regime, in which the wall moves slowly or not at all, and shorter periods of slip and micro-slip events, when the wall moves rapidly. The time axis is in units of $\delta t$, as specified in Sec.~\ref{sec:simulations}.}
    \label{fig:samples_all_measures}
\end{figure}

The analysis and detection of the slip events are based on three different measures whose potential was identified in~\cite{kramar2022intermittency}.  The first one, shown in Fig.~\ref{fig:samples_all_measures}(a),  is the horizontal velocity, $v_x(t)$,  of the top wall. The next measure is based on the properties of the differential force network; to define this network, we first discuss the concepts of the contact and force networks. 

The contact network, $\CN(t)$, is a graph with one vertex at the center of each particle (excluding those in the top and bottom walls). There is an edge between two vertices of $\CN(t)$ if the particles corresponding to these vertices are in contact at time $t$.  The force network $\FN(t)$ is obtained by assigning weights to the edges in $\CN(t)$, representing the magnitudes of the normal force acting between the corresponding particles. A differential force network, $\DFN(t)$, represents the differences between the force networks computed at two consecutive samples at $t$ and $t+\delta t$. Its edges are given by the union of the edges in the weighted graphs $\FN(t)$ and $\FN(t+\delta t)$.
The weight on an edge in $\DFN(t)$ is the absolute value of the difference of the weights assigned to this edge in $\FN(t+\delta t)$ and $\FN(t)$ if the edge is present in both. If the edge is not present in one of the force networks, we consider the corresponding weight in that force network to be zero. The measure that we use in what follows is the maximal force, $f_{plr}(t)$, for which $\DFN(t)$ percolates between the left and right boundary of the domain (in~\cite{kramar2022intermittency} top-bottom percolation was considered as well, but we do not use that measure here for brevity).

The last measure quantifies the time evolution of the force network. For each force network, $\FN(t)$, we compute its persistence diagrams $\PD_0(t)$ defined briefly here, and more fully in \cite{zomordian2004pershom}; persistence diagrams were extensively used in our earlier works in the analysis of granular matter in various settings, see, e.g.~\cite{physicaD14,pre13,pre14}.  $\PD_0(t)$ describes the structure of the supper-level sets of the force network $\FN(t)$. In particular, $\PD_0(t)$ is a finite collection of points in $(b,d)$ such that $b > d$.   Every point $(b,d)$ in $\PD_0(t)$ indicates that a  connected component appears in the set consisting of the edges of $\FN(t)$ with weights larger than $b$ and it merges with another connected component after adding all the edges with the weights larger than $d$. To help interpretation, we note that the points in $\PD_0(t)$ with $b$ coordinate sufficiently large (larger than the average force) describe the vague concept of `force chains.'  In the present work, for simplicity, we do not consider loops that could be described similarly by a separate $\PD_1(t)$, see~\cite{kramar2022intermittency} for more details.  

The difference between two persistence diagrams can be measured by their Wasserstein distance defined precisely in \cite{physicaD14}. This distance essentially sums up all the changes between the persistence diagrams in a manner analogous to the $L_2$ norm.  In what follows, we encode the changes between the force networks $\FN(t)$ and $\FN(t+\delta t)$ by considering the Wasserstein distance, $\mbox{W2B0}(t)$ between their $\PD_0$ persistence diagrams.

Figure~\ref{fig:samples_all_measures} illustrates the measures described above. At the start of the slip events detected  by the offline method, described in Sec.~\ref{sec:offline_method}, the value $v_x(t)$ is still relatively close to its baseline typical for the stick regime. This is not the case for the other two measures. In particular, $\mbox{W2B0}(t)$ increases considerably before the detection time. We exploit this behavior to develop a new online method for detecting slip events that can produce an early warning before $v_x(t)$ increases significantly. 

\subsection{\label{sec:detection}Methods for detecting slip events}

\begin{figure}
    \centering
    \includegraphics[width=0.48\textwidth]{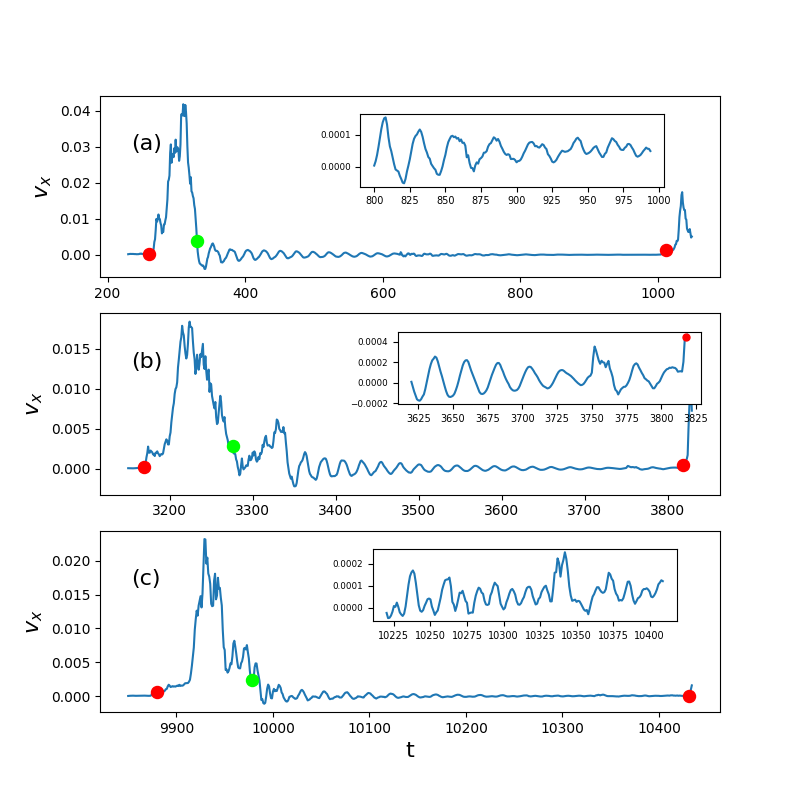}
    \caption{
    Examples of three slip events showing the magnitude of the wall velocity, $v_x$. The  red (green) dots mark the start (end) of the slip event detected by  the offline method presented in Sec.~\ref{sec:offline_method}. Note that during the stick regime, $v_x$ oscillates with decreasing amplitude around a roughly constant value.  The insets zoom into the time just before the following slip event.
    }
    \label{fig:vel_samples}
\end{figure}

Figure~\ref{fig:vel_samples} shows the stick-slip dynamics in more detail, with three illustrative examples of the behavior of the time series, $v_x$ (since there is no possibility of confusion, we drop the time dependence from now on).
Even though the peaks of $v_x$ are well pronounced during the slip events, it is non-trivial to precisely define when individual events start. Difficulties in defining the start of a slip event arise from the oscillations of $v_x$, visible in Fig.~\ref{fig:vel_samples}.  In the considered system these oscillations are related  to propagating elastic waves  caused by the vertical ($y$) component of motion of the top wall, which is associated with each slip event~\cite{kramar2022intermittency}. 

Due to these complications, slip detection must be done carefully.  One option is to employ an {\it offline} method, that uses the values of $v_x$ after the slip event has begun to distinguish it from oscillations; this approach was used in~\cite{kramar2022intermittency} and is outlined in Sec.~\ref{sec:offline_method}. Related methods were used in other works facing similar issues, see e.g.~\cite{long_granmat_2019,bares_pre17}. An alternative is to use a more sophisticated {\it online} method that accounts for the oscillations. Stochastic state space models (SSSMs), discussed later in this section, provide a good framework for such an approach. To present the main idea, we use an example of $v_x$.  Note that the amplitude of the oscillations of $v_x$ can vary and tends to decrease as we approach the onset of a slip event. Therefore, $v_x$ during  a stick can be modeled  as a composition of several components. One component represents the level around which the oscillations occur, and several harmonic components with stochastically changing amplitudes reasonably reproduce the  oscillations. To detect the start of a slip event based on  $v_x$, we exploit the fact that the behavior of this quantity is fundamentally different during the stick and slip regimes.
So, the SSSM used to model the evolution during the stick regime does not provide good predictions if the system is not in this regime. This means that during  slip events,  the difference between the value predicted by the SSSM and the observed one becomes large. Thus, to identify the start of slip events we just need to detect when the difference between the predictions and observed values  becomes sufficiently large.

Section~\ref{sec:results} demonstrates that  SSSMs can be used to devise an efficient online method for detecting the start of slip events. In particular, we will see that the algorithm tends to detect upcoming slip events earlier than the offline approach. This result opens the door for a real-time prediction of slip events. In what follows, we first review the offline approach in Sec.~\ref{sec:offline_method}, followed by a short  background of stochastic models and Kalman filter in  Sec.~\ref{sec:kalman}.

\begin{figure}
    \centering
    \includegraphics[width=0.48\textwidth]{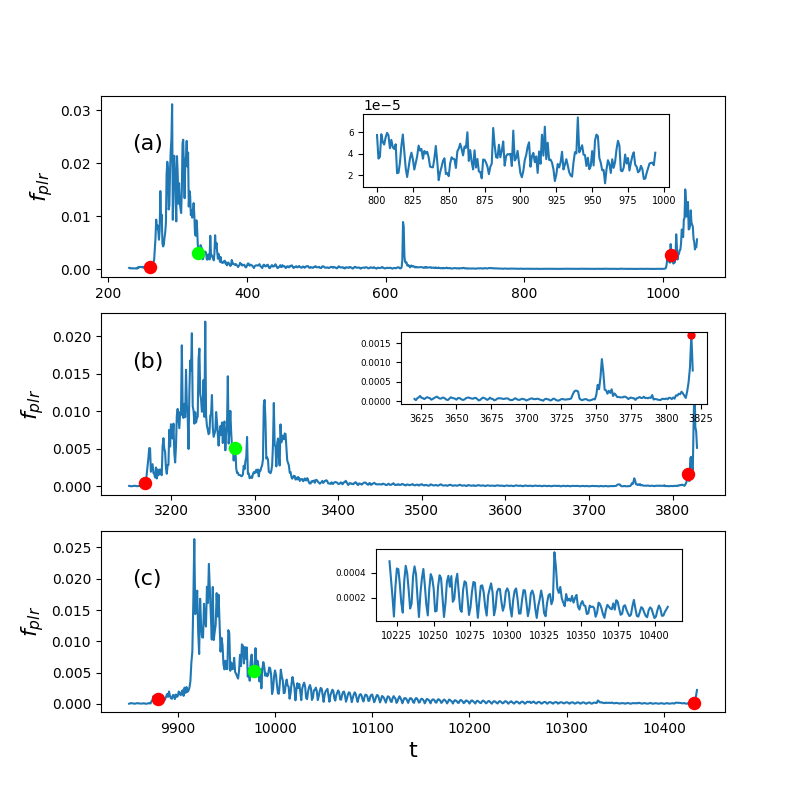}
    \caption{The percolation measure, $f_{plr}$, for the same time samples shown in Fig.~\ref{fig:vel_samples}; the meaning of the red and green dots is as discussed in the caption of Fig.~\ref{fig:vel_samples}. 
}
    \label{fig:perc_samples}
\end{figure}

\subsubsection{\label{sec:offline_method}Offline approach for detecting  slip events }

In the rest of this paper we will call slip events the events that are detected by this approach.  To deal with the oscillations present in $v_x$, shown in Fig.~\ref{fig:vel_samples}, the offline approach for detecting  slip events~\cite{kramar2022intermittency} relies on a choice of two thresholds. 
The value of the first threshold, $v_l = 2 \times 10^{-3}$, is larger than the amplitude of the  oscillations visible in the examples shown in Fig.~\ref{fig:vel_samples},  so the method avoids spurious detection of slip events. If $v_x$ exceeds $v_l$, then a slip event is identified at time $t$.  However, at this time the wall has already begun to move with a non-negligible speed.  For this reason, the offline method uses a second smaller threshold, $v_s = 2.5 \times 10^{-4}$, to find more precisely the start of each slip event.  For this purpose, the algorithm  moves back in time until the value of $v_x$ drops below $v_s$. The time when this occurs is the  starting point of the slip (the numerical values used for $v_l$ and $v_s$ are motivated  in~\cite{kramar2022intermittency}).  The end of a slip event is identified as the time at which $v_x$ as well as its average over the next 50 outputs  (roughly two periods of oscillations following a slip event) is smaller than $v_l$. This algorithm is slightly different than the one detailed in \cite{kramar2022intermittency}, which uses the condition that the average over the preceding 50 outputs marks the end of a slip event. We make this change to avoid artificial termination of a slip event, which occurs when $v_x$ drops briefly below $v_l$ at the start of a slip event.

The main shortcoming of this approach (or any other approach based on using both past and future data for slip detection), is that it cannot be used for real-time detection of slip events. It detects the time at which the event starts only after the wall velocity exceeds the specified threshold. Moreover, the precise values of the thresholds influence the number and length of the detected events.

\subsubsection{State Space Models and Kalman Filter}
\label{sec:kalman}

\begin{figure}
    \centering
    \includegraphics[width=0.48\textwidth]{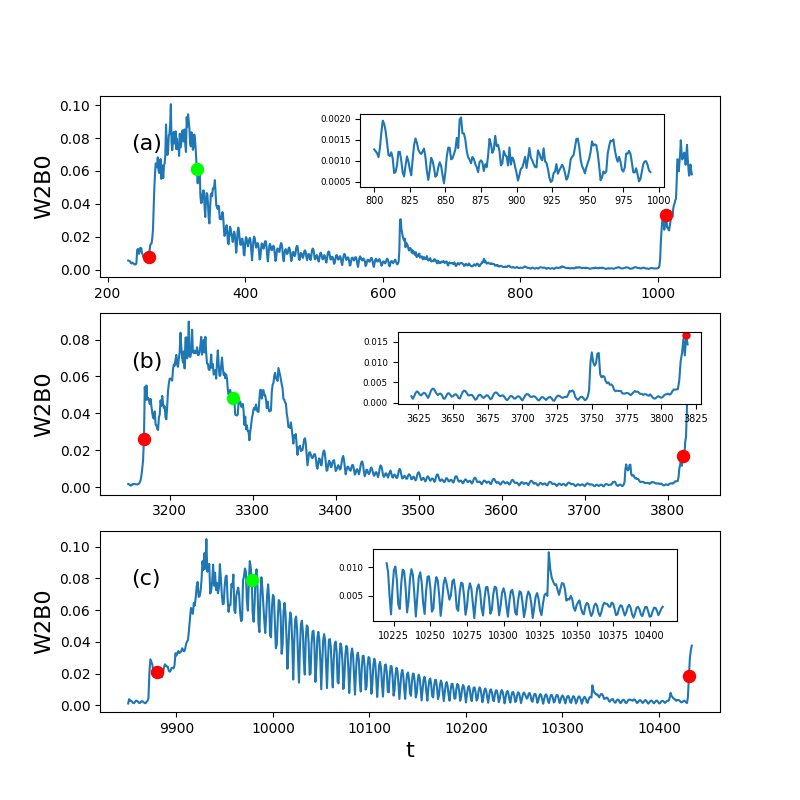}
    \caption{The W2B0 measure for the same time samples in Fig.~\ref{fig:vel_samples}. 
    Note the events at  $t=650$ in (a), $t=3750$ in (b), and $t = 10350$ in (c); these are examples of internal, local changes of the W2B0 measure that do not result in a slip event. 
    The meaning of the red and green dots is as discussed in the caption of Fig.~\ref{fig:vel_samples}.  
    }
    \label{fig:w2_samples}
\end{figure}

As discussed at the beginning of Sec.~\ref{sec:detection}, during the stick regime the signal specified by $v_x$ can be approximated by a composition of several components, one of them representing the level around which oscillations occur, and the rest describing oscillations.  The behavior of the other measures is more complex. Figs.~\ref{fig:perc_samples}~and~\ref{fig:w2_samples} show the $f_{plr}$ and W2B0 measures during the same time windows as in Fig.~\ref{fig:vel_samples}.  For example, to model W2B0, we have to account for both the oscillations and a decreasing trend of the signal.

We proceed by presenting the general idea behind stochastic modeling of a time series $y_t$,  which can be decomposed into several  components whose characteristics are captured by the value of the state vector $\bm{\theta}$, which encodes the properties such as trend and oscillations of a considered measure.   As commonly done in the literature \cite{petris2009dlm}, we use a discrete time  $t\in\mathbb N$ for the rest of this section, where $t$ is in units of $\delta t$, specified in Sec.~\ref{sec:simulations}. For example, if $y_t$ represents the magnitude of the wall velocity, $v_x$, then $y_t = v_x(t)$, the wall velocity at the physical time $t * \delta t$. To model the  time series $y_t$ by a sequence of random variables $Y_t$ we will use a special class of SSSMs known as dynamical linear models (DLMs). This choice is motivated by two independent factors. First, the necessary computations can be performed efficiently. Second, the models are sufficiently rich to properly capture the evolution of the considered measures during the stick regime while they are too rigid to assimilate  the  time evolution of these  measures  during the slip regime.  

The main assumption of the DLMs is that the evolution of the $n$-dimensional state vector, $\bm{\theta}_t$, is governed by a real Markov chain  called the state process and the random variable  $Y_t$ is a noisy  (imprecise) scalar observation of this chain. To be more precise, the evolution of $\bm{\theta}_t$ is described by the equation 
\begin{equation}
\label{eq:DLM}
    \bm{\theta}_t = G_t \bm{\theta}_{t-1} + \bm{w}_t 
\end{equation}
while the values $Y_t$ can be reconstructed from $\bm{\theta}_t$  via the equation  
\begin{equation}
\label{eq:observable}
    Y_t = F_t\bm{\theta}_t + v_t
\end{equation}
where $(G_t)_{t\geq 1}$ and $(F_t)_{t\geq 1}$ are known sequences of matrices of order $n\times n$ and $1\times n$, respectively.  Two independent sequences of random variables  describing the observation error, $(v_t)_{t\geq 1},$ and the model error, $(\bm{w})_{t\geq 1}$,  are distributed according to zero mean normal distributions  $v_t \sim \N_1(0, V_t)$ and $\bm{w}_t \sim \N_n(\bm{0}, W_t)$, where the variances $V_t$ and covariance matrices $W_t$ are given.  We will explain in  Sec.~\ref{sec:models} how to specify $G_t, F_t, W_t$, and $V_t$ for the measures considered in this paper.  Here we just note that  the $G_t$ models the evolution of the individual  components of the signal  and  $F_t$ describes how to  combine these components  to obtain the random variable $Y_t$, which models the time series $y_t$. The error (noise) terms are important because  there might be external noise in the physical system and  the relations described by $G_t$ are only approximate.

Unlike in deterministic systems, the value of $\bm{\theta}_t$ is not known precisely. Instead, the probability that $\bm{\theta}_t$ has a given value is expressed by a probability distribution  $\pi(\bm{\theta}_t)$, which in the case of DLMs is required to be a multivariate normal distribution. Given a DLM and the distribution  $\pi(\bm{\theta}_{t-1})$, we can  predict the value of $\bm{\theta}_t$ as
\begin{equation}
\label{eq:predicted_state}
    \bm{\hat{\theta}}_t = G_t \bm{\bar{\theta}}_{t-1}
\end{equation}
where $\bm{\bar{\theta}}_{t-1}$ is the mean value of $\pi(\bm{\theta}_{t-1})$. The predicted value, $\bm{\hat{\theta}}_t$,  is known as the forecast and can  be used to construct a prediction  
\begin{equation}
\label{eq:predicted_observation}
\hat{Y}_t = F_t \bm{\hat{\theta}}_t
\end{equation}
of the next value $y_t$ of the modeled time series. In particular, if we know $\pi(\bm{\theta}_{0})$, then  Eqs.~(\ref{eq:predicted_state})~and~(\ref{eq:predicted_observation}) allow us to  predict the value of $y_1$.  However,  we usually do not know the initial distribution  $\pi(\bm{\theta}_{0})$. The fact that $\pi(\bm{\theta}_{0})$ is not known can be overcome by using the iterative procedure known as the Kalman filter~\cite{petris2009dlm}, which uses the Bayesian probability to improve our knowledge about the distribution of $\bm{\theta}_{t}$ by incorporating  the  observed data $y_0, \ldots y_t$. This is done by  computing  the conditional distribution  $\pi(\bm{\theta}_{t} | Y_1 = y_1, \ldots,  Y_t = y_t )$.  In practice, if the DLM is an  appropriate model for the time series $y_t$ and the initial (prior) distribution $\pi(\bm{\theta}_{0})$ is given by a normal distribution with large variance, then after a short burn-in period the conditional distribution $\pi(\bm{\theta}_{t} | Y_1 = y_1, \ldots,  Y_t = y_t )$ describes the state $\bm{\theta}_{t}$ with reasonable accuracy, and the model provides satisfactory predictions of the future states of the time series $y_t$.

The precise definitions of  $G_t$ and $F_t$ depend on the particular measure considered and will be provided in Sec.~\ref{sec:models}. On the other hand, to quantify the  observation error and model error we use standard universal methods. Here we only provide their brief summary; detailed exposition is available~\cite{petris2009dlm}. The shape of the covariance matrix $W_t$  for the model error is  given  by the discount factor method, i.e.,
\begin{equation}
\label{eqn:W}
W_t  = \frac{1-\delta}{\delta} P_t,
\end{equation}
where $P_t$ is the conditional covariance of $\pi(G_t\bm{\theta}_{t-1}| Y_1 = y_1, \ldots,  Y_{t-1} = y_{t-1} )$ and $\delta \in (0,1]$ is the discount factor which will be specified for the individual measures in Sec.~\ref{sec:models}. The value of $\delta$ corresponds to the overall accuracy with which $G_t$ describes the state update from $t-1$ to $t$, referred to in \cite{petris2009dlm} as `model trust.' Note that if $\delta = 1$, then Eq.~(\ref{eq:predicted_state}) does not contain noise, and the evolution of $\bm{\theta}_t$ is deterministic.

The variance $V_t$  of the observation error is modeled by a time-invariant unknown random variable, $\sigma^2$,   estimated from the data at each time $t$. To be compatible with the iterative scheme defined by the Kalman filter, the variable  $\sigma^2$ has to be inverse gamma distributed. For the initial (prior) distribution $\pi(\sigma^2)$ we choose  the inverse gamma distribution $\mathcal{I}\mathcal{G}(2, 10^{-3})$. This is a `vague' prior in the Bayesian terminology; it expresses a high degree of uncertainty as to the value of the unknown $\sigma^2$. By choosing a vague prior, the distribution of $\sigma^2$ quickly responds to the data.

\subsubsection{Specification of dynamical linear models (DLMs)}
\label{sec:models}

In this section we provide  DLMs  that model  the evolution of the measures, defined in Sec.~\ref{sec:measures}, during the stick regime.  To specify a DLM for a given measure we start by specifying the components of the state vector $\bm{\theta}$ that capture the  trend and oscillations of the signal.  To describe the time evolution  of the individual components of $\bm{\theta}$,  given by the state process, Eq.~(\ref{eq:DLM}), we need to set $G_t$ and the covariance matrix $W_t$ of the model error $\bm{w}_t$. We recall that  $W_t$ is specified by choosing the discount factor $\delta$. Finally, to forecast the value of the modeled  measure, by using Eq.~(\ref{eq:observable}), we need to specify the observation function, $F_t$, and the distribution of the observation error. As explained in Sec.~\ref{sec:kalman}, the observation error is modeled by a time-invariant random variable estimated from the data at each time $t$. The required  initial distribution is always  chosen to be the inverse gamma distribution $\pi(\sigma^2) \dist \mathcal{I}\mathcal{G}(2, 10^{-3})$.

Figures~\ref{fig:vel_samples},~\ref{fig:perc_samples},~and~\ref{fig:w2_samples} manifest that  all considered  measures exhibit oscillations, albeit with different frequencies. To model the oscillations we use a few  harmonic components. For a given measure the frequency of the first harmonic component is set to its dominant frequency $\omega$, determined by Fourier analysis of the first $10\%$ of the signal. The frequencies of the other harmonic components  are natural multiples of $\omega$. So, for each model, we specify the number of harmonic components and the discount factor $\delta$.

Our extensive  numerical investigation shows that  all results discussed in Sec. \ref{sec:results} are robust under rather large  variations of these two parameters. The final choice of the parameters for each model was guided by diagnostics  of the normalized prediction error (NME) of the model  defined as
\begin{equation}
e_t = \frac{y_t - \hat{Y}_t}{\sqrt{Q_t}}, 
\label{eq:standardized_error}
\end{equation}
where $Q_t = F_t( P_t + W_t) F'_t + V_t$, $F'_t$ is the transpose of the matrix $F_t$ and $P_t$ is the conditional covariance of $\pi(G_t\bm{\theta}_{t-1}| Y_1 = y_1, \ldots,  Y_{t-1} = y_{t-1} )$. In other words, $Q_t$ is a covariance of the conditional distribution of $\pi(Y_t| Y_1 = y_1, \ldots,  Y_{t-1} = y_{t-1})$. If the model is a good fit for the data, then the NMEs are  uncorrelated and they are distributed according to the standard normal distribution~\cite{petris2009dlm}. We use the Ljung-Box statistic to test~\cite{ljung1978ljungboxtest} for possible autocorrelation and the Shapiro-Wilk test~\cite{shapiro1965shapirowilktest} to see if the normalized  errors are distributed according to  the standard normal distribution. We chose the models that yield the best diagnostics during  the stick regimes shown in  Figs.~\ref{fig:vel_samples},~\ref{fig:perc_samples},~and~\ref{fig:w2_samples}.

We start by setting up the model for the wall velocity, $v_x$.  Figure~\ref{fig:vel_samples} shows that during the stick regime $v_x$ oscillates around an almost constant value. Note that oscillations exhibit multiple frequencies.  Their dominant frequency $\omega = 0.2476$ was determined from the first $10\%$ of the data.  These oscillations can be modeled satisfactorily with only  two harmonic components with frequencies $\omega_i = i*\omega$, for $i = 1,2$.  The harmonic component with the frequency $\omega_i$ is represented  by a pair of two variables  $\theta^{2i-1}, \theta^{2i}$ that evolve in such a way that the point   $(\theta^{2i-1}_t, \theta^{2i}_t)$ rotates around a circle with the constant frequency $\omega_i$. Without the error (noise) terms, the radius of this circle would be constant as well. However, the error terms allow for a slow change of this radius. 

The value around which the oscillations occur is represented  by the last component $\theta^5$ of the state vector $\bm{\theta}$. The  value of $\theta^5_t$ changes very slowly, and  on the deterministic level we consider $\theta^5_t$ to be constant. We allow it to change only via the error term. The following equations, defining  $G_t$, formalize the described evolution of the components of $\bm{\theta}$
\begin{subequations}\label{eq:wall_model_forecast}
 \begin{align}
    \theta^{2i-1}_t &=  \theta^{2i}_{t-1} \sin\omega_i + \theta^{2i-1}_{t-1} \cos\omega_i,\\
    \theta^{2i}_t &=   \theta^{2i}_{t-1}\cos\omega_i -\theta^{2i-1}_{t-1}\sin\omega_i,\\
    \theta_t^5 &= \theta^5_{t-1},
\end{align}  
\end{subequations}
for $i = 1,2$. We note that both variables $(\theta^{2i-1}_t, \theta^{2i}_t)$, for each value $i = 1,2$, oscillate with the same frequency and amplitude. Only their phase is different. To reconstruct the oscillations around the slowly changing value $\theta^5_t$,  it is sufficient to consider the variables  $\theta^{2i-1}_t$ for $i = 1,2$. The observation function, $F_t$, just adds the individual components of the signal and is defined as 
\begin{equation}
F_t(\bm{\theta}_t) = \theta^{1}_t + \theta^{3}_t + \theta^{5}_t.
\end{equation}
To specify  the model error, $\bm{w}_t$, we use the discount factor $\delta = 0.6$. To initialize the DLM  we need to define the initial (prior) distribution for the state $\bm{\theta}_0$. We use  a normal distribution centered at mean $\bm{\bar{\theta}}_0$ where $\bar{\theta}_0^j = 0$, for $j=1, \ldots 4$,  $\bar{\theta}_0^5 = v_x(0)$, with 
the width specified by the scaled covariance matrix $\bar{C}_0 = I$.

Now we turn our attention to the left-right percolation force, $f_{plr}$, shown in Fig.~\ref{fig:perc_samples}. The dominant  frequency of oscillations for  this measure is $\omega = 0.5401$. Again it is sufficient to use two harmonic components with frequencies $\omega_i = i*\omega$,  $i = 1,2$. As in the case of $v_x$, the evolution of the variables $\theta^{1}, \ldots, \theta^4$, representing the harmonic  components, is described by Eqs.~(\ref{eq:wall_model_forecast}a)~and~(\ref{eq:wall_model_forecast}b).

The most important difference between the behavior of $v_x$ and $f_{plr}$ is that the value around which $f_{plr}$ oscillates is not changing slowly. This value decreases in an approximately linear manner. To incorporate this decrease into the model, we need two  state components: $\theta_t^5$ which models the slowly changing slope, and  $\theta_t^{6}$ which captures the  decreasing value around which the signal oscillates.  These two components evolve according to the following equations
\begin{subequations}\label{eq:perc_model_forecast}
\begin{align}
    \theta_t^{5} &= \theta_{t-1}^{5},\\
    \theta_t^{6} &= \theta^{6}_{t-1 } + \theta^{6}_{t-1}.
\end{align}
\end{subequations}
The observation function, $F_t$, again sums the variables capturing the oscillations and the variable $\theta^{6}_t$ representing the value around which the oscillations occur, so $F_t(\bm{\theta}_t) = \theta_t^1 + \theta_t^3 + \theta_t^6$. To specify  the observation error we use the discount factor $\delta = 0.93$. As in the case of $v_x$, the initial distribution for the state vector $\bm{\theta}_0$ is a normal distribution centered at $\bm{\bar{\theta}}_0$ where $\bar{\theta}_0^j = 0$, for $j=1, \ldots 5$,  $\bar{\theta}_0^{6} = f_{plr}(0)$ and  the scaled covariance matrix $\bar{C}_0 = I$.

The last considered measure, W2B0, exhibits oscillations with the dominant frequency $\omega = 0.5387$, and using two harmonic components is once again sufficient. The most important difference between the behavior of $f_{plr}$ and $\mbox{W2B0}$ is that the value around which W2B0 oscillates decays in a roughly exponential manner, see Figure~\ref{fig:w2_samples}. Thus, it can be modeled by a function $e^{\mu t}$ where the value of $\mu \leq 0$ is allowed to change slowly through the error term.  To incorporate this exponential decrease in the model, we use two  state components: $\theta_t^5$ which models the slowly changing value $\mu$, and  $\theta_t^{6}$ which captures the behavior of $e^{\mu t}$. The following equations ensure the desired evolution of these components 
\begin{subequations}\label{eq:w2_model_forecast}
\begin{align}
    \theta_t^{5} &= \theta_{t-1}^{5},\\
    \theta_t^{6} &= \theta^{6}_{t-1} e^{\theta^{5}_{t-1}}.
\end{align}
\end{subequations}
The  discount factor is $\delta = 0.67$ and  the initial distribution for the state vector $\bm{\theta}_0$ is a normal distribution centered at $\bm{\bar{\theta}}_0$ where $\bar{\theta}_0^j = 0$, for $j=1, \ldots 5$,  $\bar{\theta}_0^{6} = \mbox{W2B0}(0)$ and  the scaled covariance matrix $\bar{C}_0 = I$.

Equation~(\ref{eq:w2_model_forecast}b) is non-linear, so the model is not a DLM and  the Kalman filter  cannot be used to compute the necessary conditional distributions. To accommodate the non-linearity of the function $G_t$, we use the extended Kalman filter (EKF)~\cite{evensen2009da} which uses the Jacobian matrix of $G_t(\bm{m}_{t-1})$ to compute the covariance of the predictive distribution of $\bm{\theta}_t$.  There are well-known divergence problems with EKF~\cite{evensen2009da}, since  using the Jacobian matrix can  lead to unrealistically small values of the covariance matrix of the state distribution.  This can impede the ability of the filter to assimilate the observed data and eventually cause divergence of the model. In our case, this happens when the value of  $\theta_t^{6}$  becomes too small and is rounded to zero. To retain numerical integrity during our computations, we re-initialize the filter upon encountering filter divergence. We note that this occurred  only once in the entire data set.

\section{Results}
\label{sec:results}

Now we are ready to show how the dynamical linear models (DLMs) defined in Sec.~\ref{sec:models} can be used for the online detection of the slip events  identified by the offline method.  We will see that the DLMs detect not only the slip events but also additional events. The DLMs based on the wall velocity, $v_x$, and the measure of the global change in the force network, $f_{plr}$, detect roughly the same events. However, the global changes of the force network are typically  detected before the increased activity of the wall. The DLM based on W2B0, measuring local changes of the force network, identifies almost all  events detected by the other two DLMs, and in addition, it detects extra events in which a local change of the force network appears but does not spread across the whole system. As expected, in events detected by both W2B0 and $f_{plr}$, the local changes of the force network are detected before the global changes.

\subsection{Event detection by DLMs}

\begin{figure}
    \centering
    \includegraphics[width=0.48\textwidth]{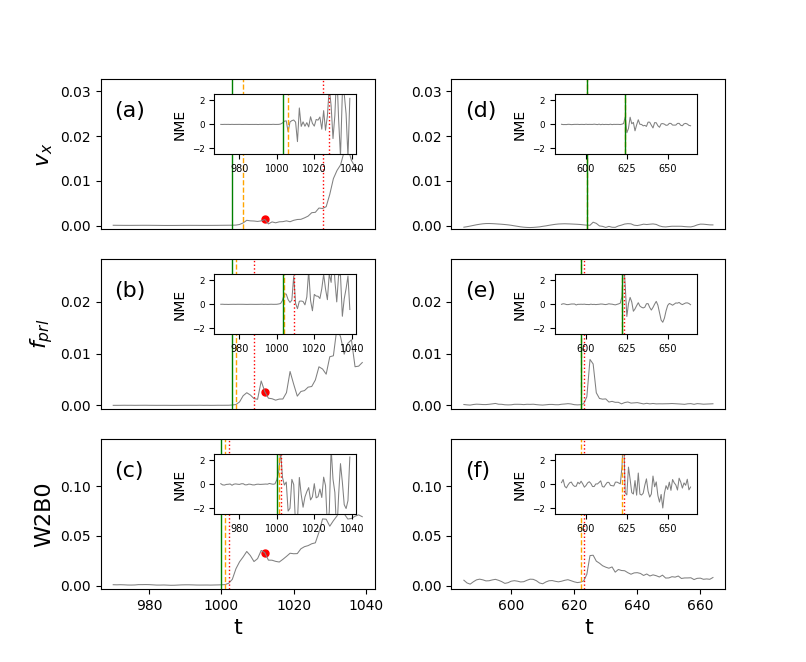}
    \caption{
    Two detection examples: (a)-(c) a slip event detected by the offline method (shown in Figs. \ref{fig:vel_samples}(b), \ref{fig:perc_samples}(b), and \ref{fig:w2_samples}(b)), and (d)-(f) a micro-slip event. The measure $v_x$ is shown in (a), (d), $f_{plr}$ in (b), (e) and W2B0 in (c), (f).  In (a)-(c), the red dots identify the start of the slip event determined by the offline method. The insets show the respective NME for individual DLMs, with the vertical lines marking detection times for individual DLMs using three different values of $T_e = 0.1,~0.4$, and $1.5$ for the solid green line, dashed orange line, and dotted red line, respectively.} 
    \label{fig:detection_examples}
\end{figure}

By construction, the accuracy of the predicted values of a measure modeled by the  DLM significantly deteriorates during the slip events detected by the offline method. To identify the events, during which  the time evolution of a considered measure does not conform to the respective DLM, we choose a maximal acceptable value of the NME, $T_e$. The events detectable by the DLM and a given error threshold, $T_e$, are obtained as follows.  The start of an event is considered to occur at time $t_0$ if the NME, $e_t$, satisfies $|e_{t_0-1}| < T_e$ and $|e_{t_0}| \geq T_e$. While one could define that the event ends when the value of $|e_t|$ drops below $T_e$, it is possible that  the value of  $|e_t|$ becomes accidentally small during a slip event detected by the offline method. Therefore, to declare  that the event ends at $t_1$, we require that all the values $|e_{t_1}|, |e_{t_1 - 1}|, \ldots, |e_{t_1-m}|$ are  smaller than $T_e$.  The value of $m$ is chosen to be roughly one period of the dominant oscillations present during the stick regime (see, e.g., Fig.~\ref{fig:vel_samples}) corresponding to $m=25$.

To illustrate the dependence of the starting time  of a detected event  on the value of $T_e$, we proceed by considering an example of two events detectable by all DLMs. The first event, shown in Figs.~\ref{fig:detection_examples}(a)-(c), is a slip event detected by the offline method.
In contrast, Figs.~\ref{fig:detection_examples}(d)-(f) show a micro-slip, which is not detected by the offline method. This nomenclature is motivated by the literature~\cite{Dalton2001_pre,Johnson:2005kt,Johnson:2007cs,ciamarra_prl10,long_granmat_2019}, which also distinguishes slips and micro-slips.  In the present work,  micro-slips are events that are not detected by the offline method but are detected by the DLM based on  $v_x$ for $T_e = 0.1$. This particular value of $T_e$ is motivated below. In what follows we will also see that there are events  detectable only by the DLM based on W2B0. Any  event detected by W2B0  for $T_e = 0.4$ which cannot be detected by the DLM for $f_{plr}$  is called a  `local change'; this particular numerical value is motivated below.  

Figures~\ref{fig:detection_examples}(a)-(c) compare detection times of a slip event obtained by  DLMs of  $v_x$, $f_{plr}$, and W2B0, respectively, for  three different values of $T_e$.  Clearly, decreasing the value of $T_e$  leads to an earlier detection time. We note that for each fixed value of $T_e$, the DLM for W2B0 detects the event first, followed by $f_{plr}$, and then $v_x$. 

Figures~\ref{fig:detection_examples}(d)-(f) show a micro-slip event  and  illustrate some difficulties that may arise  if a single  threshold value, $T_e$, is chosen. Namely, the DLM for $v_x$  does not detect this event for $T_e = 1.5$ and $T_e = 0.4$, since the NME rises only slightly above $0.1$. On the other hand, using $T_e = 0.1$ for W2B0 does not lead to a transition detection inside the depicted time window. The start of the event is  detected  before the shown time window starts and the event lasts through the whole time window.  Therefore, if  $T_e$ is too large, then  slip events may be detected late or even missed, while choosing  $T_e$ too small may result in erroneously classifying parts of a stick regime as a slip, micro-slip, or local change event. For this reason, in what follows we will consider a range of $T_e$ values. 

\begin{figure}
    \centering
    \includegraphics[width=0.48\textwidth]{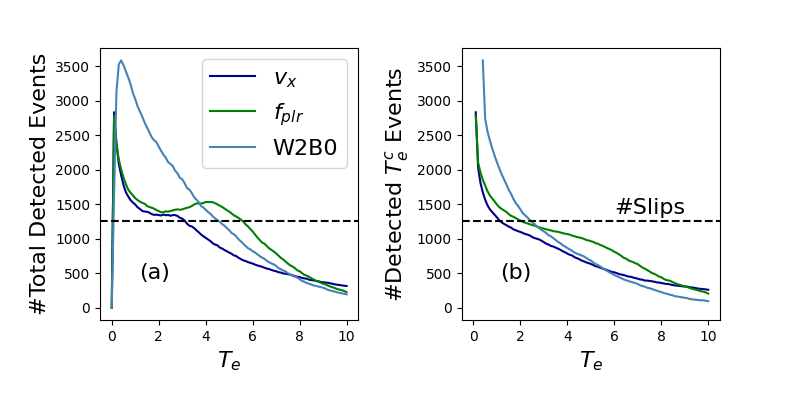}
    \caption{(a) The total number of  events detected by DLMs of $v_x$, $f_{plr}$, and W2B0 as a function of $T_e$. The curves corresponding to $v_x$ and $f_{plr}$ show a well-pronounced maximum at $T_e^c = 0.1$, while the curve corresponding to W2B0 shows a maximum at $T_e^c = 0.4$. The dashed line indicates the number of slip events detected by the offline method. (b) The number of events after removing aftershocks, i.e.,  the number of events detected at $T_e^c$ that are still detectable at $T_e$.}
    \label{fig:detection_count}
\end{figure}

With this example in mind, we now examine the entire data set. Figure~\ref{fig:detection_count}(a) plots  the number of events detected by DLMs for the considered measures. We observe that each of the measures shows a prominent peak, located at $T_e^c = 0.1$ for $v_x$ and $f_{plr}$, and $T_e^c = 0.4$ for W2B0.  If we decrease $T_e$ below $T_e^c$ (as specified for each of the measures), then distinct events start to merge and their number decreases. Eventually, as $T_e$ approaches zero, all the events merge into a single one. Thus, from now on we will only consider the values of $T_e \geq T_e^c$.  Note that earlier in this section,  we used the value of $T_e^c = 0.1$ ($T_e^c = 0.4$)  at which the largest number of events can be detected by $v_x$ (W2B0) to define  micro-slip (local change) events.

As the value of $T_e$ increases, the number of detected events decreases. This is caused by the disappearance of events during which the NME does not exceed $T_e$, see e.g., Fig.~\ref{fig:detection_examples}(d). Figure~\ref{fig:detection_count}(a) shows that 
for $v_x$ and $f_{plr}$ there is a pronounced plateau followed by a gentle rise.  Further investigation shows that some events are followed by  almost immediate aftershocks, see Fig.~\ref{fig:vel_samples}(b). Detection of these aftershocks  causes this rise because,  for small  values of $T_e$, the main event and its aftershock are identified as one event;  however, as $T_e$ increases, both the main event and the aftershock are eventually identified as separate events. To avoid the identification of aftershocks as separate events and to accurately capture the number of distinct events, we consider an adjusted number of events, shown in Fig.~\ref{fig:detection_count}(b). The  aftershocks are removed by using the events detected at $T_e^c$ as a baseline. Different  events detected at $T_e \geq T_e^c$ are counted as one if they occur during a  single event detected at $T_e^c$. The adjusted number of slip events decreases for all DLMs and values $T_e \geq T_e^c$.

\subsection{The insight provided by different measures}

We continue by investigating the relation between the events detected by DLMs for different measures at $T_e = T_e^c$.  We say that two events detected by different measures are matched if the time intervals over which they take place overlap.  By matching the events detected by $v_x$, we find that $93\%$ of these events (slips and micro-slips) are also detected by $f_{plr}$. Moreover, $97\%$ ($98\%$) of the events detected by $v_x$ ($f_{plr}$) are detected by W2B0.  Hence,  a disturbance of the wall, that causes the NME for $v_x$ to exceed $T_e^c$, is typically accompanied by a local as well as  global rearrangement of the force network detected by W2B0 and $f_{plr}$ respectively.

\begin{figure}
    \centering
    \includegraphics[width=0.48\textwidth]{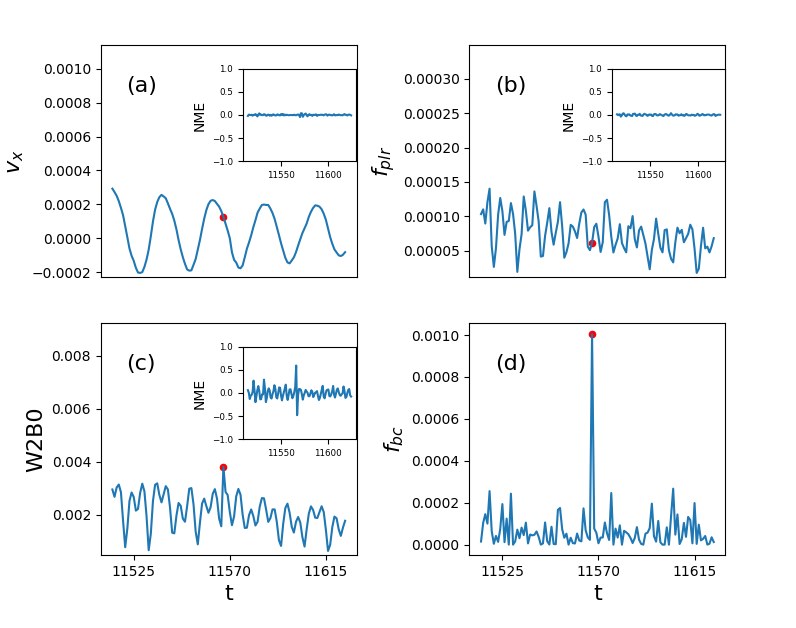}
    \caption{Example of an event detected by the W2B0 at $T_e = 0.4$ that is not detected  by $v_x$ or $f_{plr}$ at $T_e=0.1$. (a)-(c): $v_x, f_{plr}$, and W2B0 with the inset displaying the corresponding NME; (d) the broken contact force, $f_{bc}$.  The red point marks the detection made by the W2B0 at $T_e = 0.4$. The fact that $v_x$ and $f_{plr}$ do not detect this event suggests that it corresponds to a local change in the force network. The corresponding spike in $f_{bc}$  suggests that the localized change is  strongly related to broken contacts.
    }
    \label{fig:w2_b0_only}
\end{figure}

On the other hand,  $28\%$ ($26\%$) of the events detected  by W2B0 at $T_e^c$ are not detected  through $v_x$ ($f_{plr}$).  Figure~\ref{fig:w2_b0_only} shows an example of a local change event detected solely by W2B0.  Figures~\ref{fig:w2_b0_only}(a)-(b) show that for $v_x$ and $f_{plr}$, the signal does not exhibit any strong deviation from the stick behavior modeled by the DLMs, and consequently their NMEs  are small. However, the W2B0 signal significantly deviates from the predicted oscillatory behavior, see Fig.~\ref{fig:w2_b0_only}(c). The fact that  there is no visible change in the $f_{plr}$ quantifying  the global changes of the network implies that the detected change of the force network is local. For this particular local change, we find that there is a large spike in the broken force, $f_{bc}$,  defined as the sum of all (normal) forces at the contacts that are broken between time $t$ and the following output time, see Fig.~\ref{fig:w2_b0_only}(d). The spike in $f_{bc}$ appearing at the same time as the spike in W2B0 indicates that, at least for this local change event, the change in W2B0 is strongly related to broken contacts. We note that similar local changes, including non-affine motion or rotations of the particles, have been discussed  in previous works~\cite{Maloney2004_prl,bares_pre17}.

Now we turn our attention to the relation between the slip events detected by the offline method and  events detected by  the DLMs.  At each $T_e$, we match the events detected by the DLMs to the  slip events.  To better understand the composition of the matched events, we use two ratios: true positive rate and false positive rate. For a given DLM, the true positive rate, $r_{tp}(T_e)$, is the ratio of the number of  slip events that are detected by the DLM,  and the total number of  slip events.  The  false positive rate, $r_{fp}(T_e)$, is the ratio of the events detected by the DLM that do not match any slip event, to the total number of events detected by the DLM. The large  false positive rates are caused by the fact that the DLMs also detect micro-slip events and local change events which are not detected by the offline method. At this point,  we cannot reliably predict which local change events will lead to a slip event. Further research will be necessary to address this issue.

\begin{figure}
    \centering
    \includegraphics[width=0.48\textwidth]{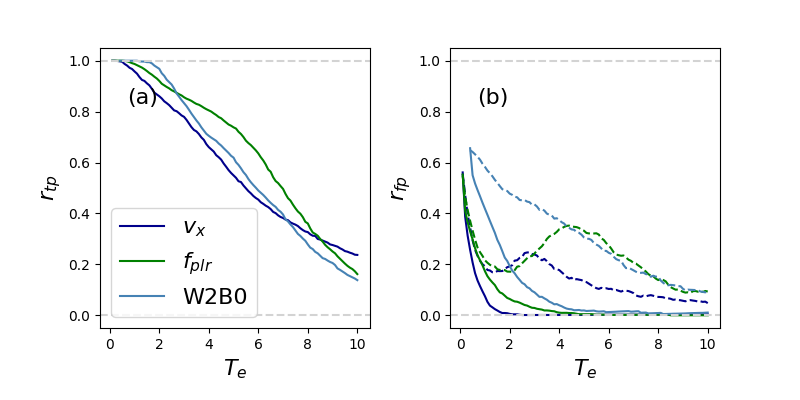}
    \caption{
        (a) True positive rate, $r_{tp}(T_e)$, and (b) false positive rates, $r_{fp}(T_e)$. The dashed line marks the false positive rate and the solid line marks the adjusted false positive rate obtained by removing the aftershocks.}
    \label{fig:accuracy}
\end{figure}

Figure~\ref{fig:accuracy} depicts the ratios, $r_{tp}(T_e)$ and $r_{fp}(T_e)$.  Part (a) 
shows that for each DLM there is a range of  values $T_e$ for which $r_{tp}(T_e) = 1$ and thus each DLM is capable of detecting all slip events.  For every DLM there is a considerable number of false positives at $T_e = T_e^c$, see Fig.~\ref{fig:accuracy}(b). Initially,  $r_{fp}(T_e)$ decreases for all the measures but then it increases for both $v_x$ and $f_{plr}$ (dashed lines). As before, this increase is caused by detecting the  aftershocks as separate events, see Fig.~\ref{fig:detection_count}. Thus, we also consider an adjusted  $r_{fp}(T_e)$ in which the aftershocks are not included. Figure~\ref{fig:accuracy}(b) shows that the adjusted $r_{fp}(T_e)$  is always decreasing and drops much faster than the non-adjusted $r_{fp}(T_e)$.

To justify our earlier classification of the events detected by $v_x$  into slips and micro-slips we return to Fig.~\ref{fig:detection_count}(b).  The slope of the curve representing the adjusted number of events detected by $v_x$ changes the fastest around the value of $T_e$ (we call it the change point) where it crosses the dashed line representing the number of slip events detected by the offline method. This is indicative of the presence of two classes of events~\cite{mori2006earthquakes, arcangelis_iop11}. Since at the change point the  number of adjusted events  is close to the number of slip events, and $r_{fp}$ for $v_x$ becomes zero (therefore the micro-slip events completely disappear),  
the two classes mentioned above directly correspond to slip and micro-slip events. That is, the distinction between the two types of events is not artificial, but reflective of an inherent difference between them.

The same argument that we  used for $v_x$ demonstrates that  the events detected by $f_{plr}$ also split into two classes roughly corresponding to slip and micro-slip events. This suggests that during slip and micro-slip events there is a close connection between the wall activity and the global rearrangements  of the force network measured by $f_{plr}$.  The  W2B0 measure does not yield a separation of events into two obvious classes  because the number of adjusted events  detected by the DLM for W2B0, see Fig.~\ref{fig:detection_count}(b), does not exhibit any clear change point.   Thus, the connection between the wall activity and local changes of the force network, measured by W2B0, is much less obvious, since local changes may or may not lead to a slip or micro-slip event.

\begin{figure}
    \centering
    \includegraphics[width=0.48\textwidth]{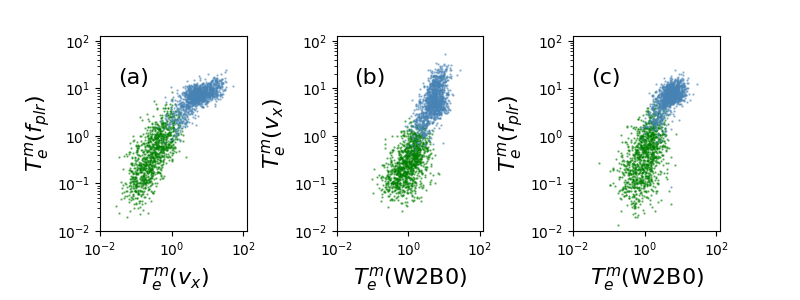}
    \caption{Scatter plot of the  $T_e^m$ values for  events detected at $T_e^c$ by both considered measures (a) $v_x$ and $f_{plr}$, (b) W2B0 and $v_x$, and (c) W2B0 and $f_{plr}$. The blue color indicates slip events while the green color indicates micro-slip events.}
    \label{fig:T_e_comparison}
\end{figure}

To further investigate the relation between wall movement and force network rearrangements, we introduce a different measure, $T_e^m> T_e$, which is defined as the largest value of $T_e$ for which an event is still detectable.  Figure~\ref{fig:T_e_comparison} shows the relation between the values of $T_e^m$ obtained by different measures, with each point corresponding to an event  detected by both measures specified on the axes labels.

For both slip and micro-slip events, Fig.~\ref{fig:T_e_comparison}(a) demonstrates   a linear relation between the magnitudes by which the horizontal  wall velocity ($v_x$) and the global change of the force network ($f_{plr}$) deviate  from the behavior expected in the stick regime. Note that the slope of the data points in this figure is different for the slip and micro-slip events. For the slip events, the wall speed increases more rapidly with $f_{plr}$. We note that if $T_e^m$ obtained by $f_{plr}$ is larger than $\approx 3$, then the event is almost certainly a slip event (see also Fig.~\ref{fig:accuracy}(b)), while the events with $T_e^m < 1$ are micro-slips.  Thus, there is a relatively well-defined critical value for the `size' of the global change of the force network, and surpassing this value leads to a slip event. Figure~\ref{fig:T_e_comparison}(b) indicates that determining whether a local change event will cause a slip event or not is more complicated. While very large (small) values of $T_e^m$ always correspond  to a slip (micro-slip) event, there is a wide range of the values $T_e^m$ obtained by W2B0 for which we observe both slip and micro-slip events.

\begin{figure}
    \centering
    \includegraphics[width=0.48\textwidth]{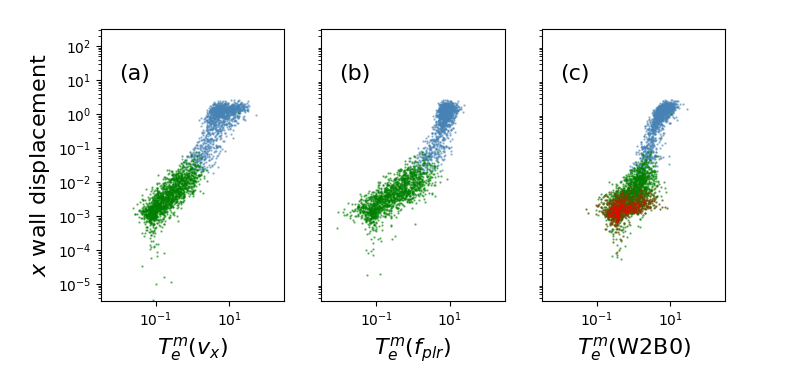}
    \caption{Scatter plot of the  wall displacement during an event and its values $T_e^m$, for events detected at $T_e^c$ by  (a) $v_x$, (b) $f_{plr}$, and (c) W2B0. The blue color indicates slip events while the green color denotes micro-slip events. In c) the red color indicates local change events.}
    \label{fig:T_e_slip_size}
\end{figure}

Next, we explore the relation between the value of $T_e^m$ and the wall displacement during a detected event. Figure~\ref{fig:T_e_slip_size}(a) shows that the values of $T_e^m$ obtained by the  $v_x$ measure are linearly proportional to the wall displacement.  We also observe a relatively clear distinction between micro-slip and slip events, since the data in  Fig.~\ref{fig:T_e_slip_size}(b) shows a clear change  of slope  around the boundary between the slip and micro-slip events. This change further indicates that for the slip events the wall activity increases more rapidly with the growing global change in the force network, than for the micro-slips. Figure~\ref{fig:T_e_slip_size}(c) relates the size of the local change of the force  network and the wall displacement.  Similarly to Figs.~\ref{fig:T_e_comparison}(b)-(c), there is a large range of values $T_e^m$ obtained by  W2B0 for which both slip and micro-slip events can occur. 
A local change of the force network with  $T_e^m$ in this range  has uncertain  consequences. It might  dissipate without causing  a global change in the force network or  trigger a global change accompanied by a micro-slip or  slip event. However, sufficiently small local changes never trigger  a slip event  while sufficiently large ones always do.

\subsection{Predictive power of different measures}
\label{sec:predictive}

So far, we have compared the events detected by individual measures. It is also of interest to examine how early the slip events, identified by the offline method, can be detected by  DLMs.  Figures~\ref{fig:detection_examples}(a)-(c) suggest that the time at which the start of a slip event is detected by a DLM depends on both the considered measure and $T_e$. To examine this quantitatively, we consider the  times at which the offline method detects the start of the slip events as a baseline and compare them with the  times obtained by different measures and values of $T_e$.  We emphasize that the purpose of this comparison is quantifying the effectiveness of the online methods based on DLMs, and not comparing the performance of online and offline methods. The offline approach uses future data for identifying slip events and is not appropriate for predictive purposes.

\begin{figure}
    \centering
    \includegraphics[width=0.48\textwidth]{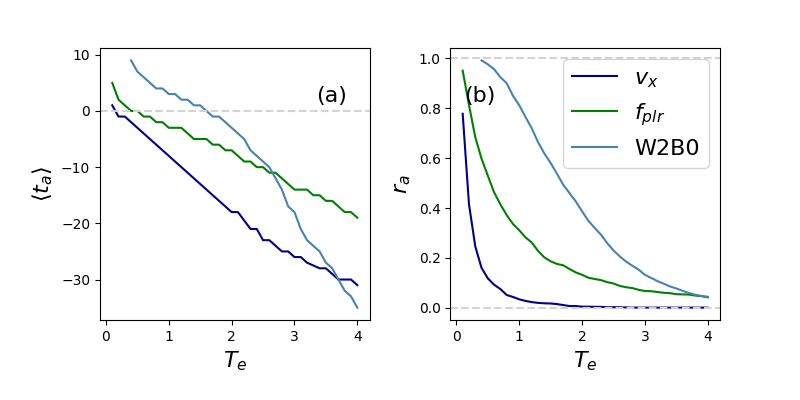}
    \caption{Comparison of offline and online methods: a) the median advance notice $\langle t_a \rangle$  taken over  slip events detected by the given measure at $T_e$; (b) the ratio of  slip events detected in advance $r_a(T_e)$.}
    \label{fig:accuracy_advance}
\end{figure}

For a given slip event, let $t_0$ and $t'_0(T_e)$ be the times at which it is detected by the offline method, and by the DLM for  the considered  measure, respectively. We then define $t_a(T_e) =  t_0 - t'_0(T_e)$. If $t_a(T_e) > 0$, then  the slip event is identified in advance. To analyze how much in advance the slip events can be detected, using different measures and values of $T_e$, we consider the median advance notice $\langle t_a \rangle$ defined as follows. For a given measure and $T_e$, the value  $\langle t_a \rangle(T_e)$ is the median over all slip events detected by the DLM. Figure~\ref{fig:accuracy_advance}(a) shows this metric as a function of $T_e \geq T_e^c$. We immediately observe that the $W2B0$ measure tends to provide the earliest detection times for a wide range of $T_e$'s. In particular, for $T_e = T_e^c$, the median advanced notice is over ten time units, while the best value that can be achieved by $f_{plr}$ is only around five, and  $v_x$ barely detects slip events in advance at all.

To obtain the portion of the slip events that are detected in advance, we divide their number by the number of all slip events. This fraction is denoted  by $r_a(T_e)$. Figure~\ref{fig:accuracy_advance}(b) shows that $r_a$, as expected,  decreases with $T_e$ for each considered measure. Once again, the shape of the W2B0 curve is different from the other two. This curve  decreases slower and  is the only one that approaches unity as $T_e$ gets close to $T_e^c$. Hence the detection method based on this measure is the only one that can predict almost all slip events in advance for $T_e \leq 0.9$.

\begin{figure}
    \centering
    \includegraphics[width=0.4\textwidth]{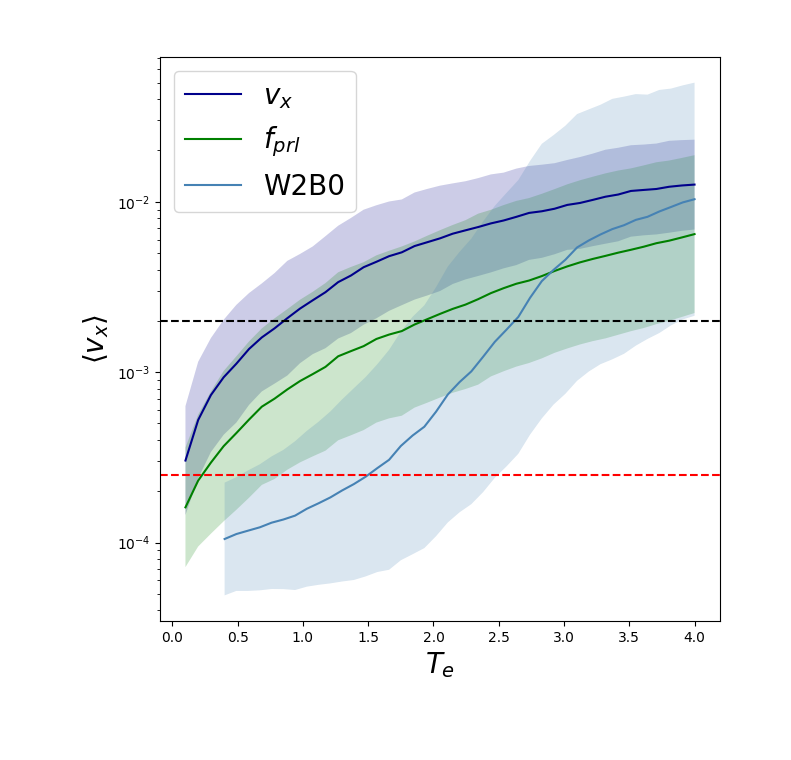}
    \caption{Median magnitude of the wall velocity at the start of slip events detected at $T_e$ by different measures.  For each measure, the solid lines mark the median of $v_x$,  and the shaded regions mark one standard deviation. The dashed lines show the two thresholds used by the offline method: (black) $v_l$ used to identify slip events and (red) $v_s$ used to identify their start.}
    \label{fig:vel_at_detection}
\end{figure}

Finally, we investigate how fast the wall moves at the  start of   the slip event detected  by different measures. Figure~\ref{fig:vel_at_detection} shows the median of $v_x$ at detection times as a function of $T_e$. The median at each fixed $T_e$ value is taken over all slip events  detectable  by the given model and $T_e$.  As expected, all the curves are increasing with $T_e$. Consistently with the results shown in Fig.~\ref{fig:accuracy_advance}, we find significantly lower values of $\langle v_x \rangle$ when considering the W2B0 measure. That is, the  detection based on the W2B0 measure takes place before the wall has begun to move appreciably.

\section{Conclusions}
\label{sec:conclusions}

In this work we present a method for detecting the onset of slip events that can be used on-the-fly for an incoming stream of data. The method is based on the fact that the behavior of the system during slip events is very different from the one during the stick regime. In particular, we consider three different measures describing the system: the wall velocity $v_x$, maximum percolation force of the differential force network, $f_{plr}$, and the Wasserstein distance between the persistence diagrams quantifying the force networks, W2B0.  For each of these measures, we build a dynamical linear model (DLM) capable of accurately predicting its behavior during the stick regime. However, the predictions provided by DLMs become inaccurate when the measures start to exhibit more complex behavior as the system approaches a slip event.   Hence, to detect the upcoming slip event we analyze the differences between the predicted and observed values of these measures.

Our analysis based on $v_x$  shows a clear distinction between slip and micro-slip events. We find that global changes in the network measured by $f_{plr}$ are essentially always associated with a slip or a micro-slip event that follows them. In particular,  the measure  $f_{plr}$ provides a slightly earlier prediction of the upcoming slip events than $v_x$. By using the W2B0 measure we identify local changes of the force network which might or might not spread over the force network and become global. We observe that only the local changes that become global are followed by a slip or micro-slip event.   While micro-slips were reported in the previous works considering similar systems, see e.g.~\cite{dahmen_group15,arcangelis_iop11}, and local changes (called `local avalanches') were discussed recently as well~\cite{bares_pre17}, we are not aware of such events being used for predictive purposes. From the predictive perspective, we note that the W2B0 measure shows the best performance, in terms of sensitivity. On the other hand, the high sensitivity of this measure also leads to a significant percentage of `false positives,' in the form of local changes that do not lead to slip events.

In summary, we observe the following timeline of the changes leading to a slip event.
Typical slip events start with a local change in the force network. The size of this local change can be quantified by the amount by which the observed values of W2B0 measure deviate from the predicted values.  This deviation roughly corresponds to an increase in the W2B0 measure. If the size of the local change is sufficiently large, then it always becomes global and triggers a slip event.  On the other hand, a sufficiently small local change does not trigger  a slip event.  There is, however, a large range of local change sizes for which the outcome is uncertain; the initial local disruption may either dissipate or trigger a global change resulting in a micro-slip or even slip event.  While the chronology of a slip described above has been discussed in the literature already, our method allows for precise quantification of this timeline.

As noted above, a significant fraction of the local changes of the force network fall into an intermediary range and may or may not result in a slip or micro-slip event. It would be very much of interest to explore whether there are distinguishing characteristics of local changes that could be used to predict if the given change will lead to a slip event.  In this work, we use W2B0 to identify local changes. This measure only compares the topological properties of the force network encoded by the persistence diagrams.  Hence precise geometry of the force network is not considered. Further research will be necessary to establish if the geometry of the force and/or contact networks can be used to predict the outcome of local changes. Our results suggest that the use of stochastic state space models and in particular DLMs could be a productive path in this direction.

\bibliography{references,granulates}

\end{document}